\documentclass[doublecol]{epl2} 
\usepackage{graphicx}
\usepackage{dcolumn}
\usepackage{bm}
\usepackage{subfigure}
\usepackage{lineno}
\usepackage{setspace}
\usepackage{amsmath}
\title{Crossover from quasi-static to dense flow regime in compressed frictional granular media.}
\shorttitle{} 
\author{F. Gimbert\inst{1,3} \and D. Amitrano \inst{1} \and J. Weiss\inst{2}}
\shortauthor{F. Gimbert \etal}
\institute{                    
  \inst{1} Institut des Sciences de la Terre, CNRS-Universit\'e Joseph Fourier, Grenoble, FRANCE, 1381 rue de la Piscine, BP 53, 38041 Grenoble Cedex 9 \\
  \inst{2} Laboratoire de Glaciologie et de G\'eophysique de l'Environnement, CNRS-Universit\'e Joseph Fourier, Grenoble, FRANCE, 54 rue Moliere, BP 96, F-38402 Saint-Martin d'H\`eres Cedex \\
  \inst{3} Now at Seismological Laboratory, California Institute of Technology, 1200 E. California Blvd., Pasadena, CA 91125, USA
 }
\abstract{We investigate the evolution of multi-scale mechanical properties towards the macroscopic mechanical instability in frictional granular media under multiaxial compressive loading. 
Spatial correlations of shear stress redistribution following nucleating contact sliding events and shear strain localization are investigated. 
We report growing correlation lengths associated to both shear stress and shear strain fields that diverge simultaneously as approaching the transition to a dense flow regime. This shows that the transition from quasi static to dense flow regime can be interpreted as a critical phase transition. Our results suggest that no shear band with a characteristic thickness has formed at the onset of instability.}

\begin{document}

\maketitle
\section{Introduction}
The mechanical behavior of granular materials is of wide concern, from natural hazard in geological context to engineering applications. However, the evolution of properties towards the flowing instability is still partially understood.

In case of packing of non-frictional, hard (non-deformable), spherical particles loaded under shear, force chains, i.e. heterogeneous distributions of contact forces on a scale much larger than the typical particle size, control the mechanical response of the granular assembly~\cite{Cates1998}. For these systems, the concept of jamming~\cite{NatureLiu1998} provides a powerful framework to analyze the onset of granular flows. These assemblies of non frictional particles exhibit jammed states resisting small stresses without irreversible deformation, whereas unjammed systems flow under any applied shear~\cite{Combe2000}. The jamming transition for such spheres at zero stress occurs at a critical value of the packing fraction $\phi$~\cite{NatureLiu1998}.

We investigate here a different situation, considering elastic (i.e. non hard) frictional disks loaded under multiaxial compression. This situation is relevant when studying geophysical instabilities (e.g. granular gouges within fault zones, landslides,...) and differs from classical configurations used to study the jamming transition~\cite{NatureLiu1998} mainly in two ways. First, considering assemblies of frictional grains, 
the parameter that controls whether the grain assembly behaves as a jammed or an unjammed state is the fraction of non-rattler grains, i.e. the fraction of grains that carry forces, rather than density~\cite{NatureBehringer2011}. Secondly, 
instead of shearing the sample at constant volume, the compressive loading conditions here considered imply a confining pressure that prevents the non-rattler fraction to evolve freely. As a consequence and contrary to the study of~\cite{NatureBehringer2011}, a percolating strong force newtork remains in the flowing phase, called the dense flow regime~\cite{CRPhysiquePouliquen2002}.

In this letter, we investigate the transition from a quasi-static regime, i.e. a regime where the sample resists to the applied stress by deforming infinitly slowly, towards a dense flow regime, where inertia comes into play. 


\section{Loading mode}
\begin{figure}
 \centering
\includegraphics[width=0.2\textwidth]{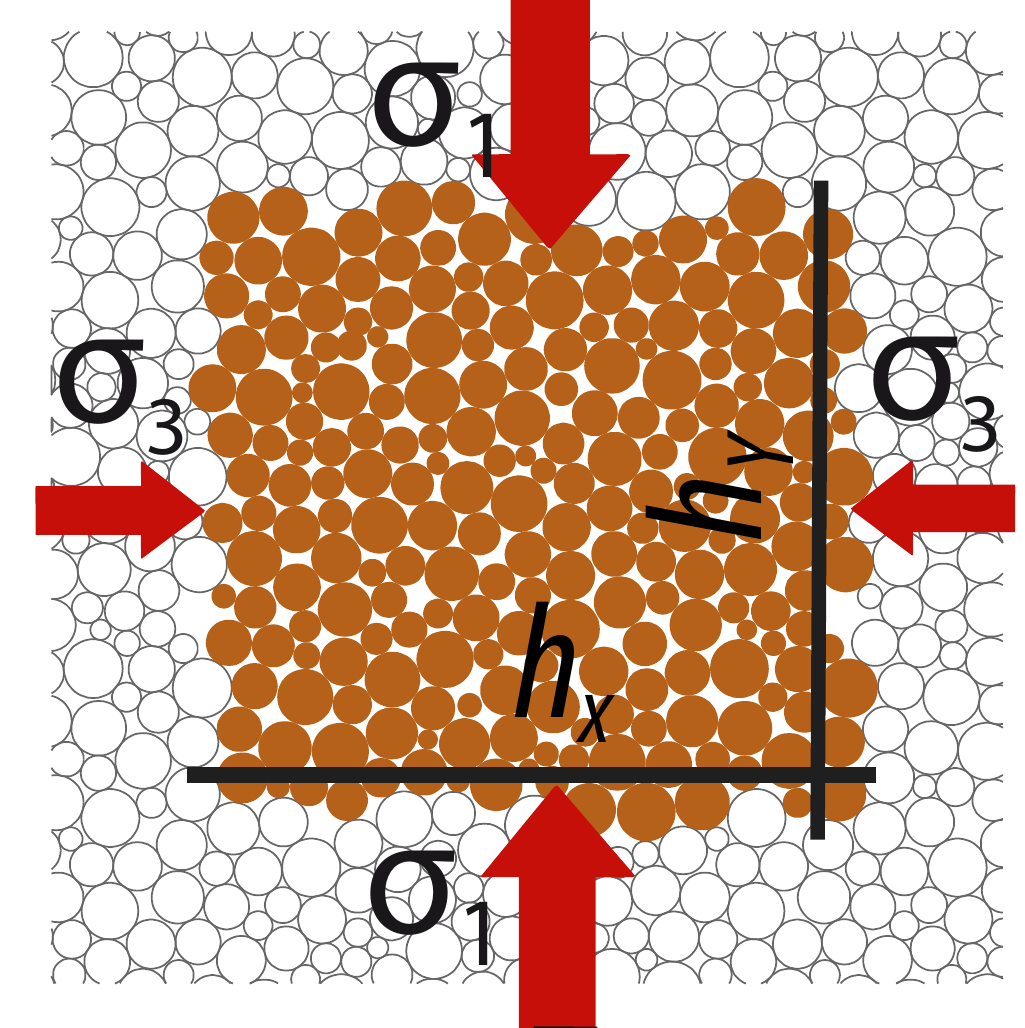}%
\caption{Illustration of the multi axial loading configuration on a sample made of 225 grains (filled circles). The sample has been replicated in all directions (unfilled circles).\label{fig::Illustration}}
\end{figure}

We consider 2D compression tests  under multiaxial loading: the axial stress $\sigma_1$ is increased whereas the radial stress $\sigma_3$, i.e. the confining pressure, is kept constant (see Figure~\ref{fig::Illustration}). Like this, the sample is sheared by increasing the deviatoric stress $\tau=\sigma_1-\sigma_3$.

\section{Position of the problem}

Laboratory experiments have been conducted with this multiaxial configuration on either continuous rocks~\cite{Mogi1967,Haimson2000} and discrete materials such as sand~\cite{Viggiani2004} or synthetic analogous materials~\cite{Hall2010}.
In continuous materials, the macroscopic instability as been first theoretically tackled by the use of the bifurcation theory, which considers a transition from an homogeneous to an heterogeneous deformation field materialized by the creation of a perennial macroscopic shear band presenting a characteristic value of thickness and spanning the whole sample~\cite{RudnickiRice1975,HaimsonRudnicki2010}. Within granular materials, computing the deformation field over a large macroscopic strain window, those perennial shear bands appear and seem to show characteristic sizes either in experiments~\cite{Viggiani2004} and simulations~\cite{Tordesillas2007}. However, this vision is counterbalanced by the heteregenous and long range correlated kinematics of quasi-static granular flow~\cite{PRLRajai2002}: 
at which temporal and spatial scales and at which stage of the loading does the granular assembly deform homogeneously? What are the relevant key features of the stress and strain fields associated to the onset of macroscopic instability? Here we investigate this problem through numerical simulations.

\section{Simulation approach}
Our simulations use the Molecular Dynamics discrete element method~\cite{BookRajaiDubois2011}. Two-dimensional granular assemblies of a number $N_g$ of frictional circular grains are considered. To characterize sample size effects~\cite{SuppMat}, we performed 320 simulations with $N_g = 2500$, 80 simulations with $N_g = 10000$ and 20 simulations with $N_g = 45000$. The results presented here concern 10000 grains samples.
The grains areas are uniformly distributed, setting the largest grain diameter $D_{max}$ such that $D_{max} = 3D_{min}$. 

The dynamic equations are solved for each grain, which interact via linear elastic laws and Coulomb friction when they are in contact~\cite{Cundall1979}. The normal contact force $f_n$ is related to the normal apparent interpenetration $\delta$ of the contacts as $f_n = k_n \times \delta$, where $k_n$ is the normal contact stiffness coefficient. The tangential component $f_t$ of the contact force is proportional to the tangential elastic relative displacement, with a tangential stiffness coefficient $k_t$. We set $k_t=k_n$. Neither cohesion between grains, nor rolling resistance is considered. The Coulomb condition $|f_t|\le \mu_{micro} f_n$, where $\mu_{micro}$ is the grain friction coefficient, requires an incremental evaluation of $f_t$ every time step, which leads to some amount of slip each time one of the equalities $f_t = \pm \mu_{micro} f_n$ is reached. A normal viscous component opposing the relative normal motion of any pair of grains in contact is also added to the elastic force $f_n$ to obtain a damping of the dynamics. 

An isotropic compression of dilute frictionless grains sets builds dense and highly coordinated initial packings of density $\phi_i \approx 0.85$ and backbone coordination number, i.e. coordination number computed over grains that carry forces~\cite{Agnolin2007}, $z_i^*~=~2N_c/(N_g(1-x_0))~=~4$, where $N_c$ is the total number of contacts and $x_0$ the fraction of rattlers grains. Then, multiaxial compression tests are performed setting the particle friction to $\mu_{micro}=1$.  

The external mechanical loading is prescribed on the grain assembly using periodic boundary conditions.
A periodic simulation cell of period $\textbf{h}$ (see section 6.3.3 of~\cite{BookRajaiDubois2011}) is considered. As the simulation cell is rectangular, the linear operator $\textbf{h}$ can be written as $\textbf{h} =\begin{pmatrix}h_x&0\\0&h_y\\\end{pmatrix}$, where $h_x$ and $h_y$ correspond to the size of the cell period in the radial and axial direction (see Figure~\ref{fig::Illustration}). Then, stresses are prescribed solving the dynamic equations of motion of $\textbf{h}$, i.e. ensuring that the internal stress computed over the whole grain assembly following equation~\ref{eq::InternalStress} (see below) counterbalances the prescribed external stresses $\sigma_1$ and $\sigma_3$. 

The axial stress $\sigma_1$ is increased at constant rate by imposing a stress increment $\delta \sigma_1^{t_r}$ at each discretisation time interval $t_r = \sqrt{\frac{m_{min}}{k_n}}/25$, where $m_{min}$ is the mass of the lightest grain. This stress control loading mode avoids stress relaxations and associated feedbacks that would be obtained under strain controlled loading, i.e. adjusting $\delta \sigma_1^{t_r}$ in order to axialy deform at constant rate $\dot{\epsilon}_1$, where $\epsilon_1$ is the axial deformation. In this stress controlled case, stress and strain localization structures develop freely.
The confining pressure $\sigma_3$ is kept constant and sized by setting the contact stiffness $\kappa = k_n/\sigma_3$ equal to 1000~\cite{BookRajaiDubois2011}. This value for $\kappa$ allows to treat elasticity in grain contacts during reasonable computational times while considering a relatively low level of deformability of grains that is relevent for application of geomechanics and geophysics. As examples, compression experiments performed on assemblies of glass beads of approximate Young's modulus $E~=~70$~GPa submitted to 100~$kPa$ of confining pressure lead to a $\kappa$-value of 700 while, from the knowledge of wave speed velocities within a granitic Earth's crust~\cite{Brocher2008}, values of $\kappa \approx 1.10^{3}$ are expected between hundreds of meters to several kilometers depth. 


 \begin{figure}
 \centering
 \quad
  \includegraphics[width=0.52\linewidth]{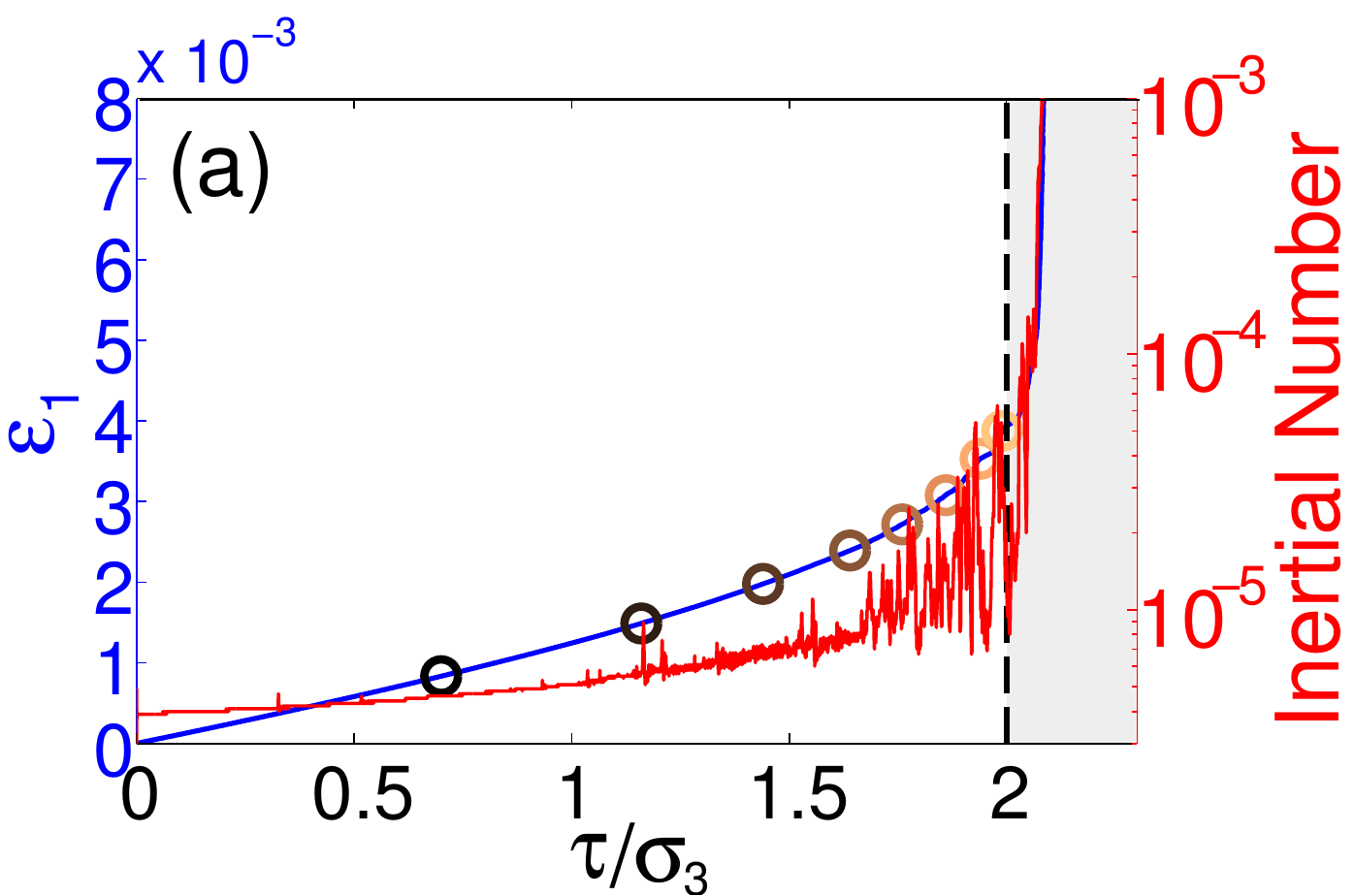}%
 \includegraphics[width=0.48\linewidth]{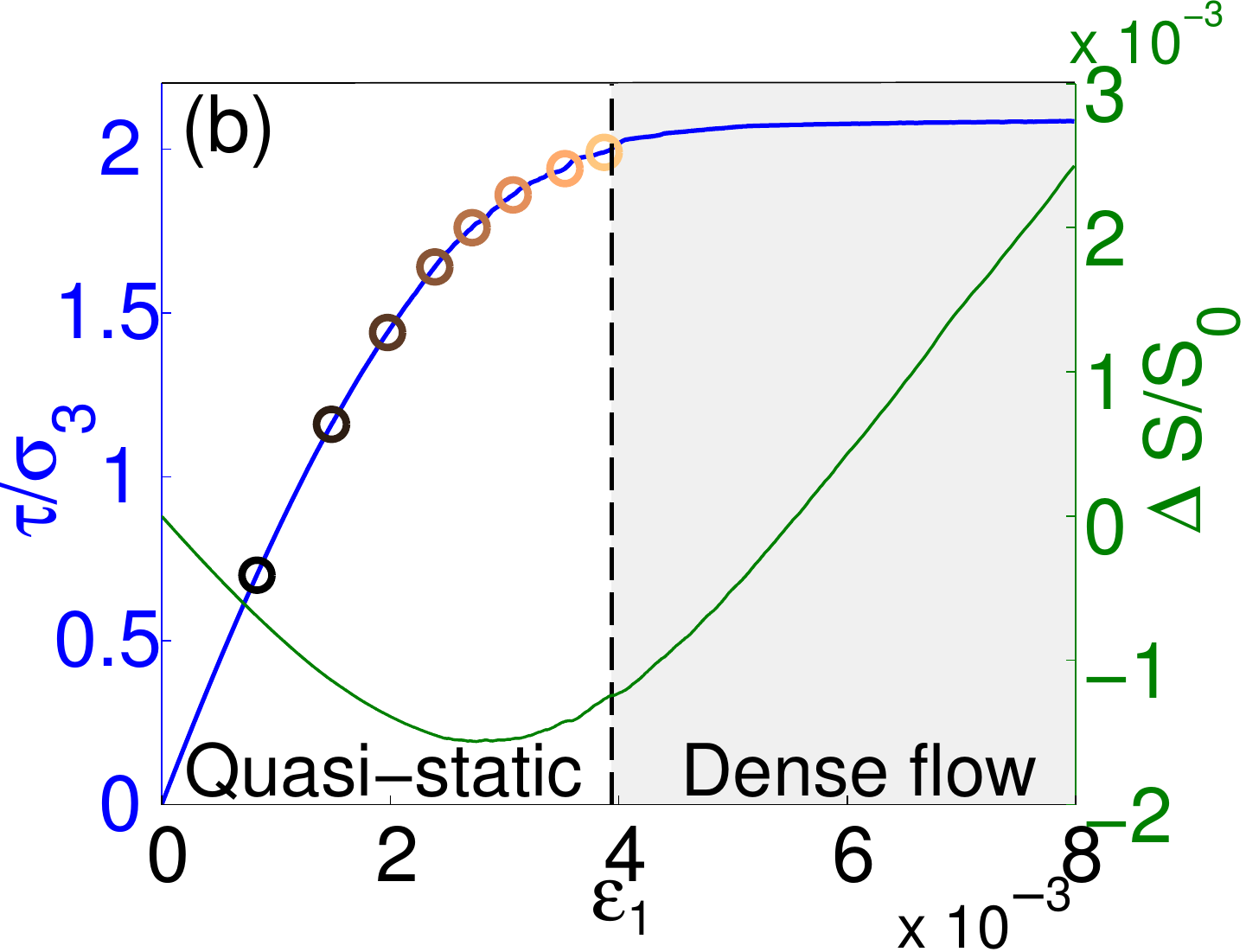}%
 \caption{Evolution of macroscopic parameters during compressional testing for a sample of 10000 grains. 
Color dots correspond to color lines on Figures~\ref{fig::StressScaleAnal}(Top) and~\ref{fig::StrainScaleAnal}(Top). Dashed lines materialize the limit between quasi-static and dense flow deformation regimes. \label{fig::MacroCurve}}
 \end{figure}
 
\section{Results}
\subsection{Macroscopic behaviour}
Figure~\ref{fig::MacroCurve} shows the macroscopic response of a granular sample loaded using $\delta \sigma_1^{t_r} = 1.10^{-6} \sigma_3$. To characterize the dynamical behaviour of the 
granular packing, we compute the inertial number $I$, which corresponds to the ratio between inertial forces and imposed forces and is defined as $I = \dot{\epsilon}_1 \sqrt{\overline{m}/\sigma_3}$~\cite{BookRajaiDubois2011}, where $\overline{m}$ is the average grain mass. Initially, $I$ is of the order of $10^{-6}-10^{-5}$ (see Figure~\ref{fig::MacroCurve}(a)). Then, when increasing $\tau$ towards $2\sigma_3$, while undergoing brutal fluctuations associated to large plastic events, $I$ remains lower than 10$^{-4}$, which is often considered as the upper bound for quasi-static conditions~\cite{BookRajaiDubois2011,GDRMidi2005}. Hence, in the region delimited by $\tau = 0$ and $\tau \approx 2\sigma_3$, the sample undergoes quasi-static deformation. At values of $\tau$ larger than $\tau_c \approx 2\sigma_3$, a brutal increase of $I$ of several orders of magnitude is observed, reaching values of the order of $10^{-3}-10^{-2}$. This indicates the transition towards a dense flow regime, where inertia comes into play. This transition is also marked when looking at $\epsilon_1$ versus $\tau/\sigma_3$, where we can see that a drastic change of slope of the curve operates around $\tau_c$ (Figure~\ref{fig::MacroCurve}(b)). For values of $\tau$ larger than $\tau_c$, the prescribed axial stress increment $\delta \sigma_1^{t_r}$ induces a large amount of axial deformation.

Figure~\ref{fig::MacroCurve}(b) also shows the surface variation $\Delta S/S_0=\frac{(h_x h_y)_{\epsilon_1}-(h_x h_y)_{0}}{(h_x h_y)_{0}}$ of the granular assembly as a function of $\epsilon_1$, where $(h_x h_y)_{\epsilon_1}$ corresponds to the sample surface computed at a given value of axial deformation $\epsilon_1$ and $(h_x h_y)_{0}$ corresponds to the initial sample surface. We observe an initial contracting phase materialized by the decrease of $\Delta S/S_0$ until a peak of contraction is reached, after which the sample dilates continuously. This contracting phase results from elastic contacts and would no longer be observed in the limit of infinitely rigid grains, i.e. infinitly large values of $\kappa$~\cite{Combe2003}. 
However, in this case, the dense flow transition is observed at negative values of $\Delta S/S_0$, i.e. at a value of packing fraction larger than the initial one. 

Thus, the transition to dense flow regime is observed when $\tau$ reaches a critical value $\tau_c \approx 2\sigma_3$, i.e. at a macroscopic friction $\mu_{macro}\sim0.5$. At this transition, stress and strain concentrations resulting from cooperative effects are expected, triggering preferential weak zones where flow is favoured. In the case of our samples made of circular grains with no rolling resistance at grains contacts, this leads to a softening of the whole granular assembly and thus to a macroscopic friction $\mu_{macro}$ much smaller than $\mu_{micro}$. According to this, studying the spatial structure of both stress and strain fields is a key point to understand the mechanisms that generate the macroscopic instability. We thus focus, in this study, on the response of the granular assembly to a small stress increment in terms of associated stress concentration and strain localization structures that form during mechanical loading.

\begin{figure}
\centering
 \begin{minipage}{0.3\linewidth}
    \begin{minipage}{0.85\linewidth}
      \includegraphics[width=1\textwidth]{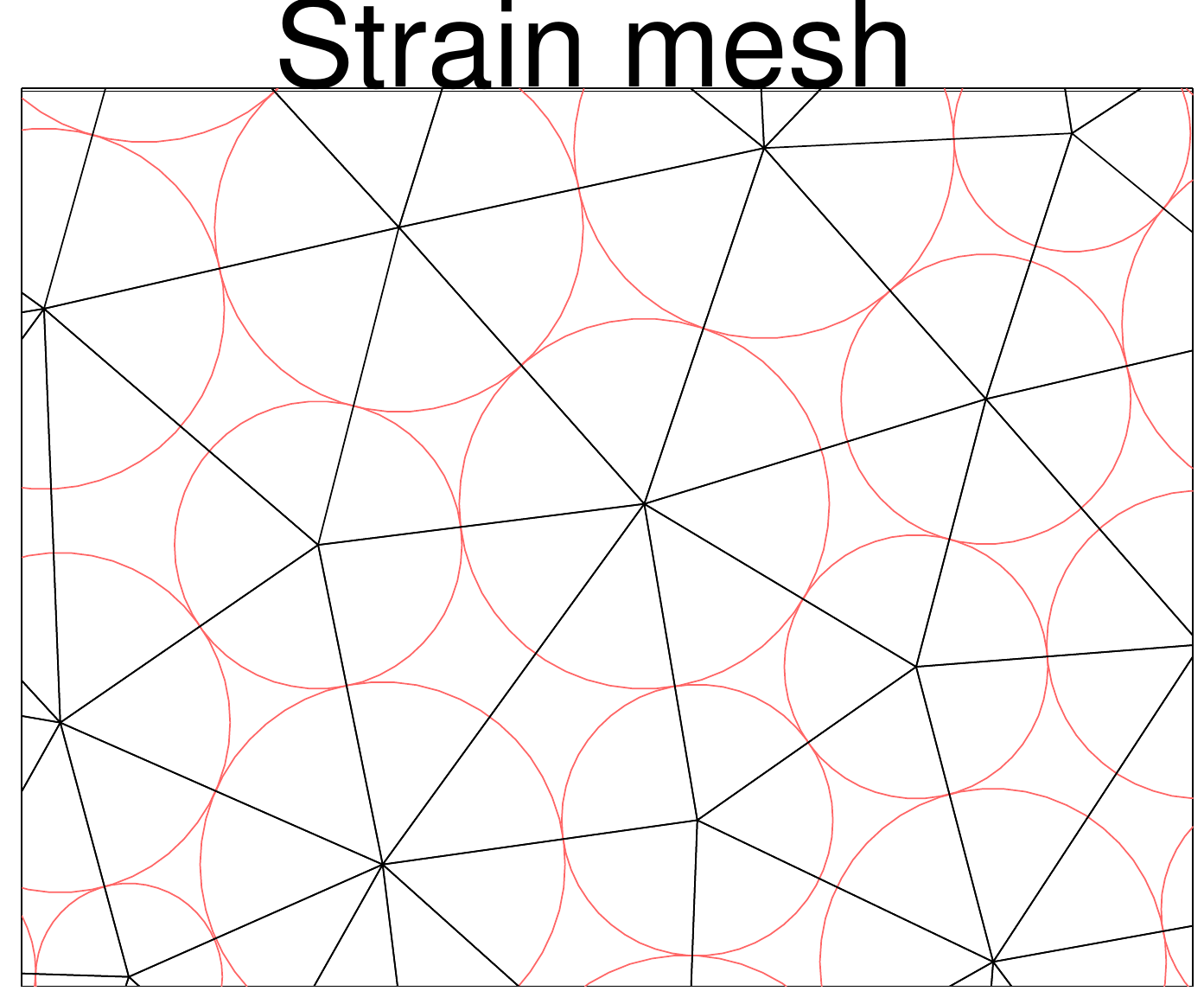}
    \end{minipage}
    \begin{minipage}{0.85\linewidth}
      \includegraphics[width=1\textwidth]{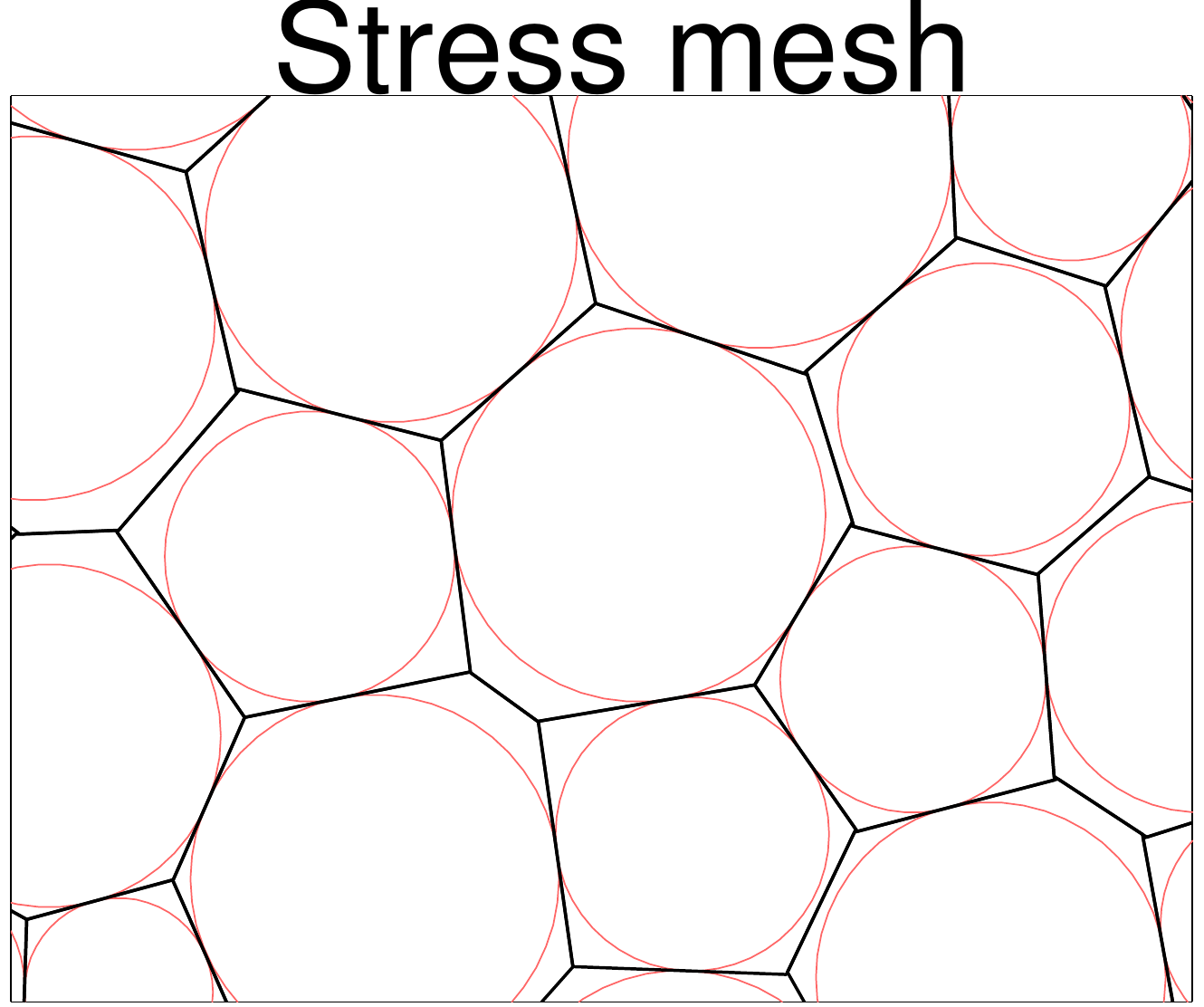}
    \end{minipage}
  \end{minipage}
 \begin{minipage}{0.45\linewidth}
 \includegraphics[width=1\textwidth]{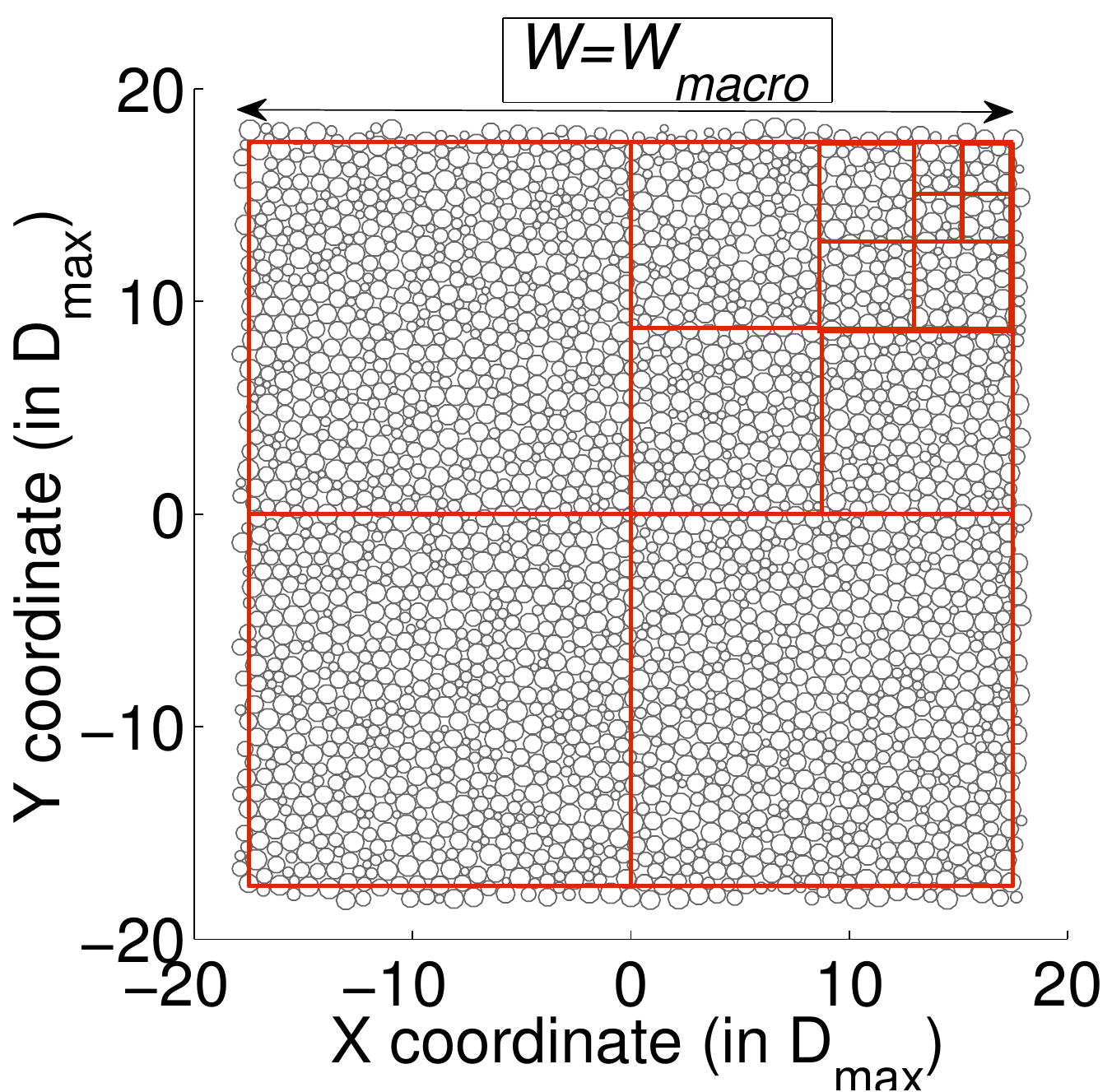}
 \end{minipage}
 \caption{Left:~Delaunay (top) and modified Voronoi (bottom) tesselations for a polydisperse granular material. Right:~Coarse graining analysis on a 2500 grains sample. \label{fig::CoarseGrainingMethod}}
 \end{figure}
\subsection{Multi-scale analysis}
We first characterize the spatial extent of regions of stress concentration by means of a coarse graining analysis~\cite{PRLMarsan2004,JStatGirard2010}: an averaged shear stress rate $<\dot{\tau}>$~\cite{Invarients} is computed at different stages of mechanical testing (cf color dots on Figure~\ref{fig::MacroCurve}) over a time window $T$ and over a broad range of spatial scales $L$, from the micro-scale corresponding to the scale of the mesh element, to the macroscale corresponding to sample size. Subsystems of the granular assembly are selected by means of square boxes of size $W$ (see Figure~\ref{fig::CoarseGrainingMethod}) and the average scale $L$ is computed as $L = \frac{1}{N_{box}} \sum_{k=1}^{N_{box}} L_k$,
where $N_{box}$ is the number of boxes of size $W$ and $L_k$ is the scale associated to the box number $k$, computed as the square root of the sum of mesh element surfaces. The standard deviation of $L_k$-values is maximum for the smallest subsystem of average size $L = 1.1 D_{max}$, and corresponds to $0.2 D_{max}$ when considering the Voronoi triangulation used to compute stresses and to $0.13 D_{max}$ when considering the Delaunay triangulation used to compute deformations (see Figure~\ref{fig::CoarseGrainingMethod}). We checked that no overlap of scales $L_k$ occurs through successive values of $W$. 

For a given assembly of grains lying within the box number $k$, the stress tensor is computed as in \cite{DeGiuliMcElwaine2011} writing 
\begin{equation}
 \sigma_{ij}^k = \frac{1}{s_k} \sum_{g_k} \sum_{c_k} (r_i^{c_k} - r_i^{g_k}) f_j^{g_k c_k}
 \label{eq::InternalStress}
\end{equation}
where $s_k = L_k^2$ is the surface associated to the grain assembly, $f_j^{g_k c_k}$ is the $j^{th}$ component of the contact force exerted on grain $g_k$ at contact $c_k$, $r_i^{c_k}$ is the $i^{th}$ component of the position vector of $c_k$, and $r_i^{g_k}$ is the $i^{th}$ component of the position vector of the center of mass of the grain $g_k$.
Considering two successive configurations, the stress rate tensor is obtained by differentiating the respective stress tensor components in time. The time resolution used is $T = \sqrt{N_g} \times 100 \times t_r$, which corresponds to the travel time of elastic waves through the granular assembly~\cite{SuppMat}.
\begin{figure}
\centering
\begin{minipage}{0.15\linewidth}
 \includegraphics[width=1\textwidth]{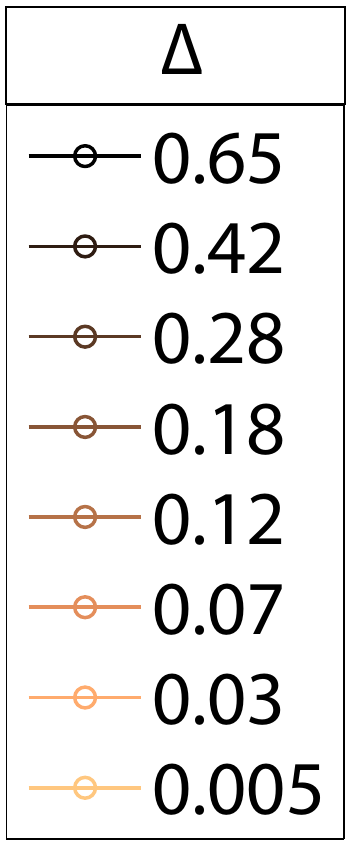}
\end{minipage}
 \begin{minipage}{0.55\linewidth}
 \includegraphics[width=1\textwidth]{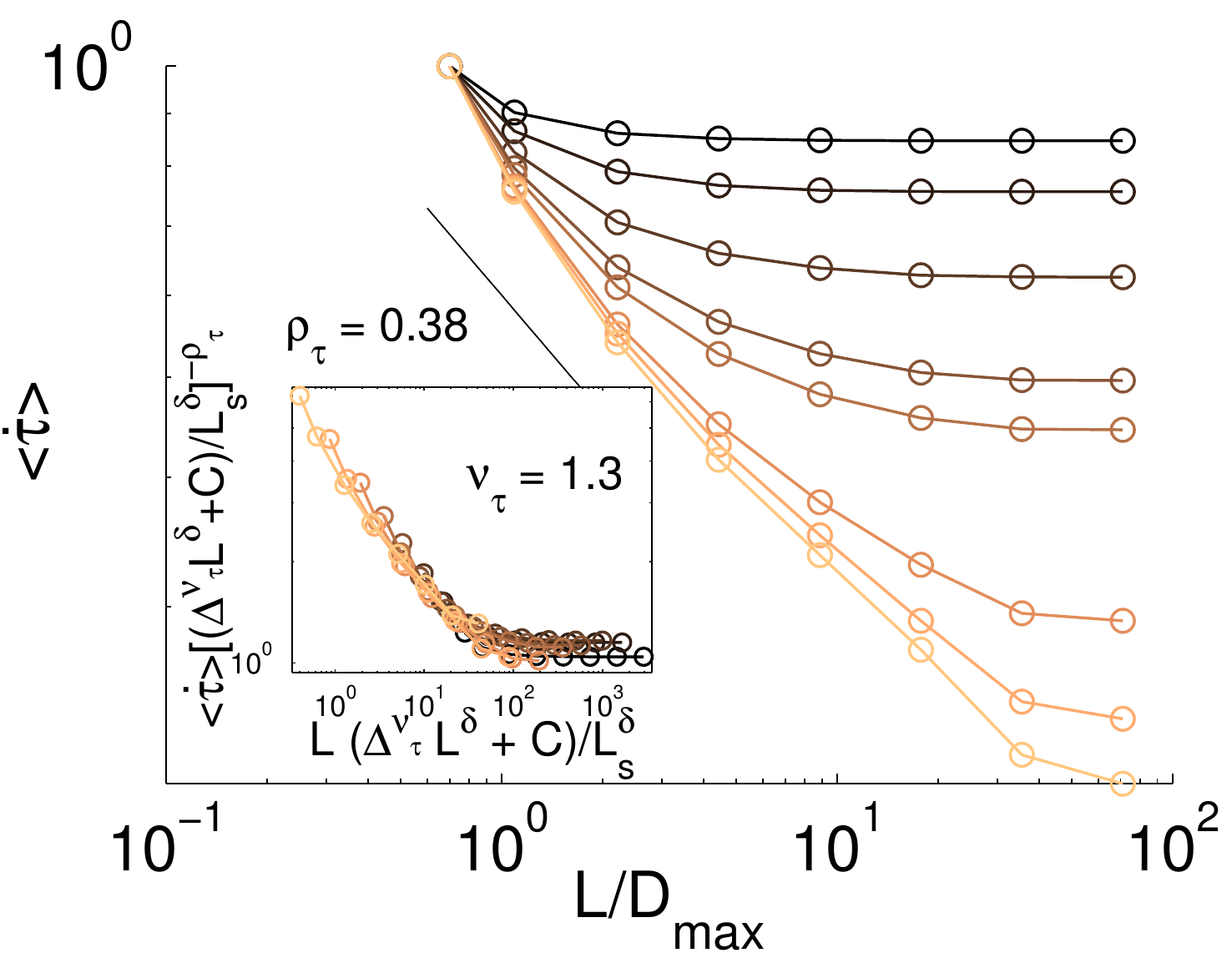}
 \end{minipage}
 \begin{minipage}{0.18\linewidth}
   \includegraphics[width=1\textwidth]{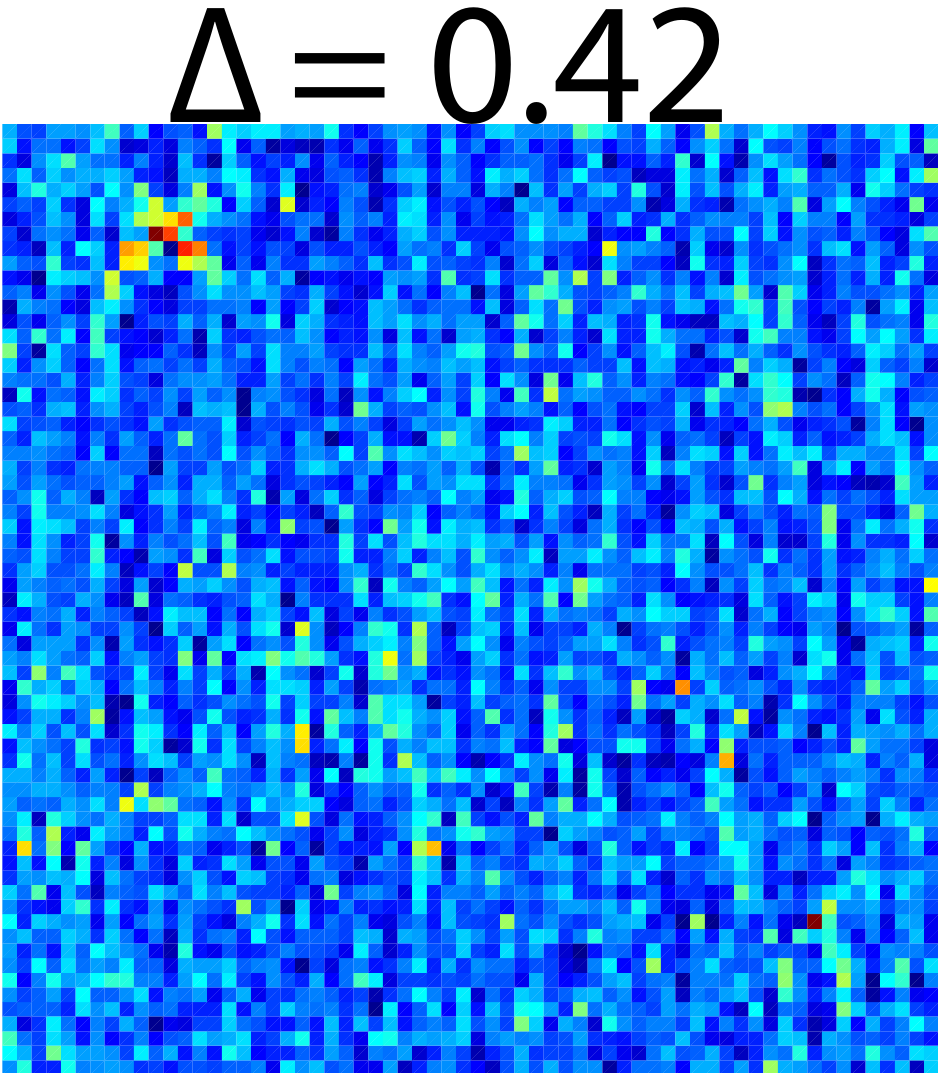}
   \includegraphics[width=1\textwidth]{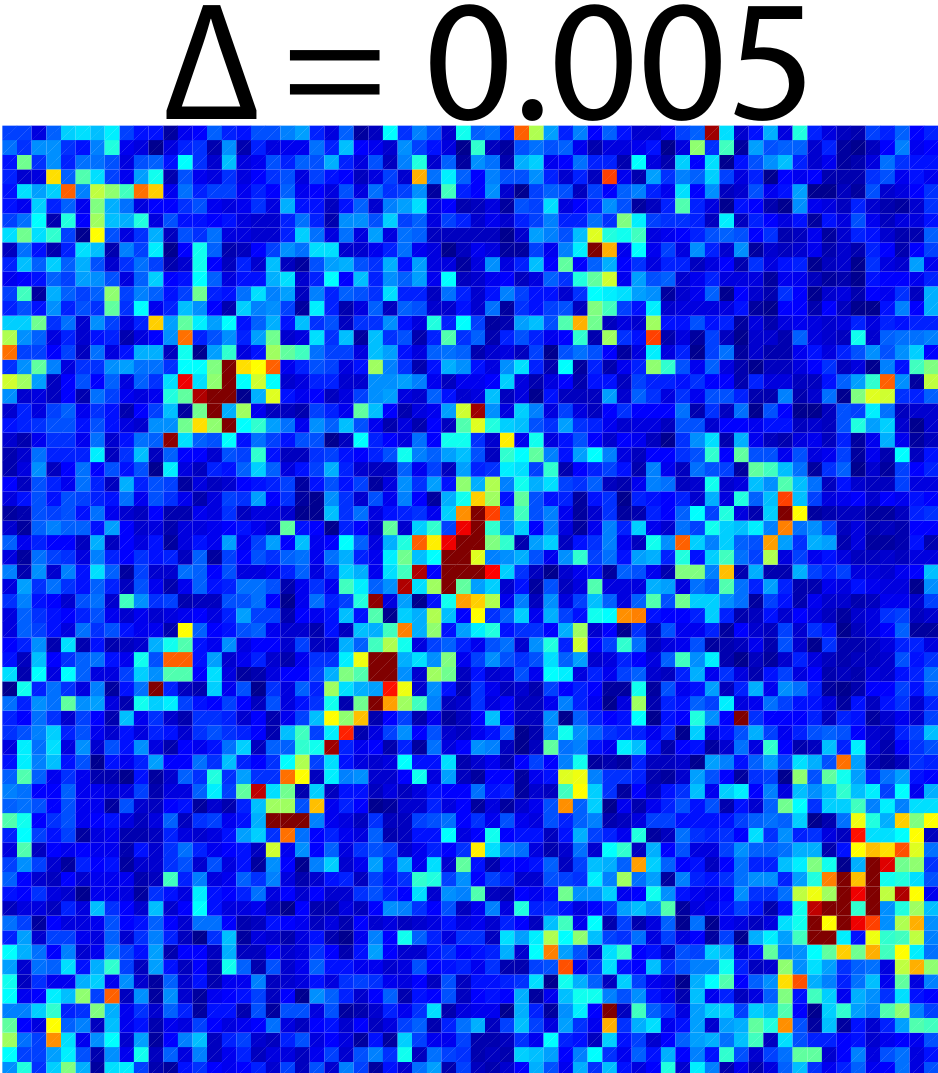}   
 \end{minipage}
 \hrule
 \begin{minipage}{0.15\linewidth}
 \includegraphics[width=1\textwidth]{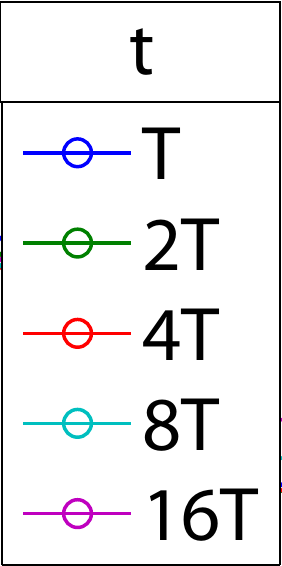}
\end{minipage}
 \begin{minipage}{0.55\linewidth}
 \includegraphics[width=1\textwidth]{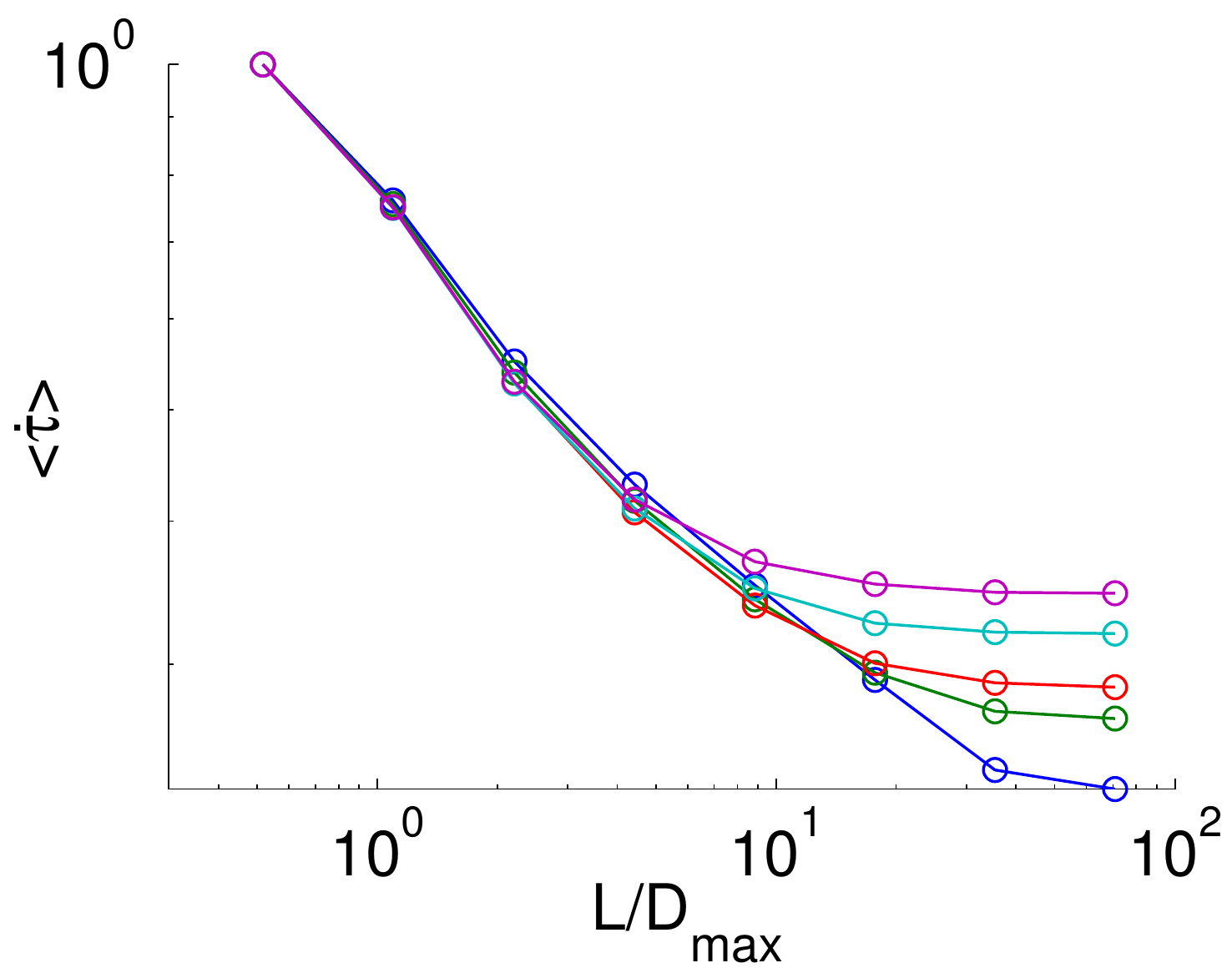}
 \end{minipage}
 \begin{minipage}{0.18\linewidth}
 \includegraphics[width=1\textwidth]{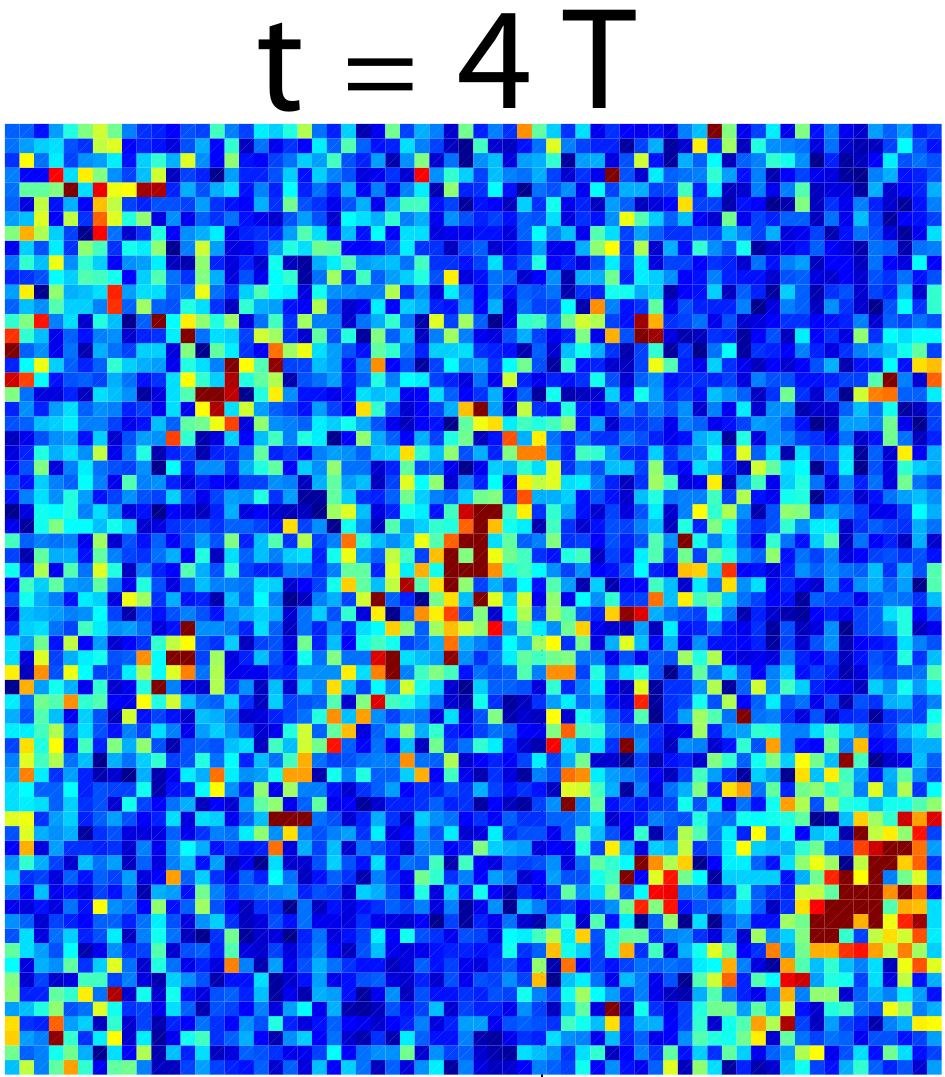}
 \includegraphics[width=1\textwidth]{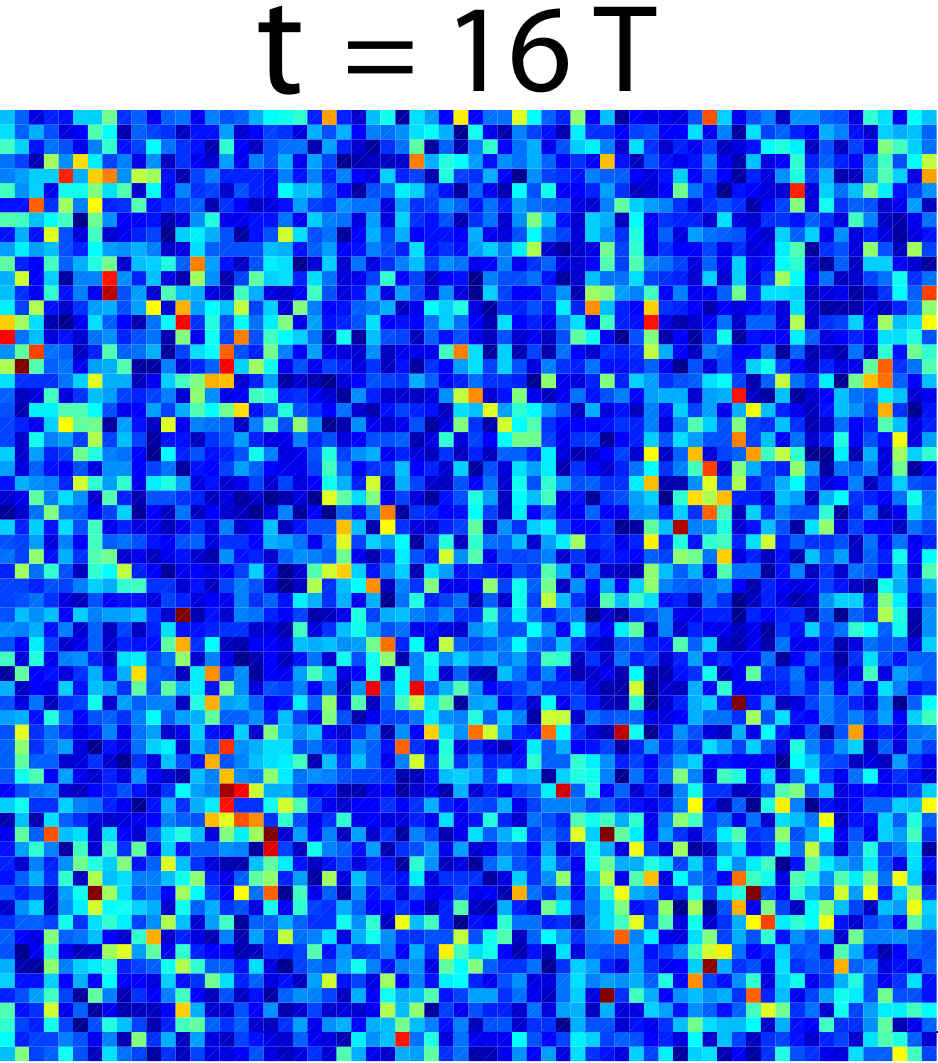}   
 \end{minipage}
\caption{Multi-scale analysis performed on the shear stress rate field $\dot{\tau}$. A selection of corresponding fields is shown on the right side: an arbitrary color scale (not shown) has been chosen for each snapshot. Top:~$<\dot{\tau}>$ versus $L$ for decreasing values of $\Delta$. Configuration locations on the stress-strain curve are shown on Figure~\ref{fig::MacroCurve}. The timescale $T$ is used to compute values of $<\dot{\tau}>$. The inset displays data collapse with respect to $\Delta$ (equation~\ref{eq::Collapse}). We find $\nu_\tau = 1.3$, $C = 0.5$ and $\delta = 1$. 
Bottom:~$<\dot{\tau}>$ versus $L$ at $\Delta = 0.005$ when increasing the timescale $t$ from $t=T$ to $t = 16T$. For graphical convenience, all computed values have been normalized by the ones computed at the micro-scale. \label{fig::StressScaleAnal}}
 \end{figure}
 
Results are shown on Figure~\ref{fig::StressScaleAnal} (Top), from the early stages of biaxial testing up to $\tau_c \approx 2\sigma_3$. The curves are selected with respect to the control parameter $\Delta$ defined as $\Delta = \frac{\tau_c - \tau}{\tau_c}$, i.e. $\Delta$ decreases as approaching $\tau_c$.
At the early stages of macroscopic deformation, a decrease of $<\dot \tau>$ with $L$ is observed at small scales while for $L$-values larger than a crossover scale $l^*_\tau$ a plateau is observed. This means that shear stress rate fields are heterogeneous for $l << l^*_\tau$, and homogeneous for $l>>l^*_\tau$. Examples of associated fields are provided on the right hand side of Figure~\ref{fig::StressScaleAnal}: at $\Delta = 0.42$, the computed snapshot shows a roughly homogeneous stress field at large spatial scales, i.e. scales larger than $l^*_\tau \approx 3 D_{max}$ in that case (see crossover scale observed on the corresponding curve), while the region of stress localization observed on the top-left part of the snapshot exhibits a spatial extension of the order of $l^*_\tau \approx 3 D_{max}$. We interpret $l^*_\tau$ as the associated correlation length~\cite{JStatGirard2010}. As macroscopic deformation proceeds, $l^*_\tau$ grows until reaching the entire size of the system at $\tau_c$ where a power law scaling $<\dot \tau> \sim L^{-\rho_\tau}$ is observed, with $\rho_\tau = 0.38$. 
The cut-off remaining on the scaling at $\Delta \rightarrow 0$ is a finite size effect~\cite{SuppMat}. At that stage, large and strongly localized structures characterize the shear stress field (see snapshot computed at $\Delta = 0.005$).
These results suggest a progressive structuring of the stress field as approaching the transition to the dense flow regime, associated to the divergence of the correlation length $l^*_\tau$. It can be verified from a collapse analysis (inset of Figure~\ref{fig::StressScaleAnal}(Top)) that $ l^* = l^*_\tau$ diverges as
\begin{equation}
l^* \sim \frac{L_s^{\delta}}{\Delta^{\nu} L_s^{\delta} + C}
 \label{eq::Collapse}
\end{equation}
where $L_s$ is the square root of the sample area and $\nu = \nu_\tau = 1.3 \pm 0.1$ is the exponent of divergence. Parameters $\delta$ and $C$ characterize the finite size effect~\cite{SuppMat}. 
A similar analysis performed on other moments $<\dot \tau^q>$ of the shear rate~\cite{SuppMat} confirms the divergence of $l_\tau^* \sim \Delta^{-\nu_\tau}$ at $\Delta~\rightarrow~0$, and reveals the multi-fractality of the shear rate field at the critical point.
These particular features of the shear stress field are observed only at the specific timescale $t = T$ corresponding to the travel time of elastic waves. This multi-scale behaviour is no longer observed at larger timescales (Figure~\ref{fig::StressScaleAnal}(Bottom)), as a clear departure from power law is observed for $t > T$, associated to the progressive homogenization of the corresponding fields, i.e. decrease of $l^*_\tau$, as $t$ increases (see also associated snapshots). Hence, the multi-scale properties of the shear stress rate field are only observed at the time scale corresponding to the time of propagation of the elastic information throughout the sample. Beyond this time, a loss of scaling properties is observed, explained by the superposition of several uncorrelated events in time, consistant with a spatially correlated stress structure associated with little memory, limited to the travel time of an elastic wave.


\begin{figure}
\centering
\begin{minipage}{0.15\linewidth}
 \includegraphics[width=1\textwidth]{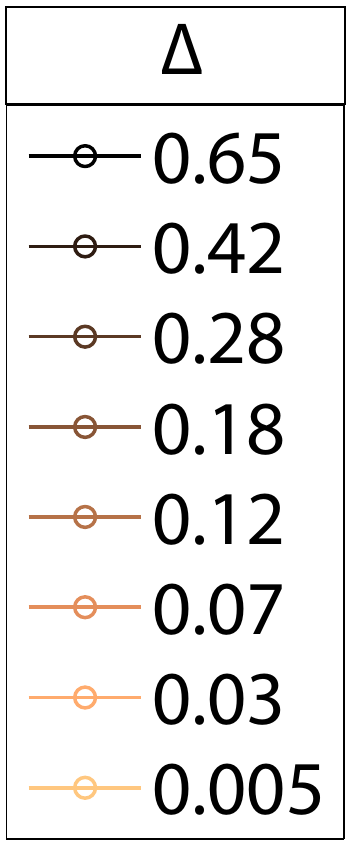}
\end{minipage}
 \begin{minipage}{0.55\linewidth}
 \includegraphics[width=1\textwidth]{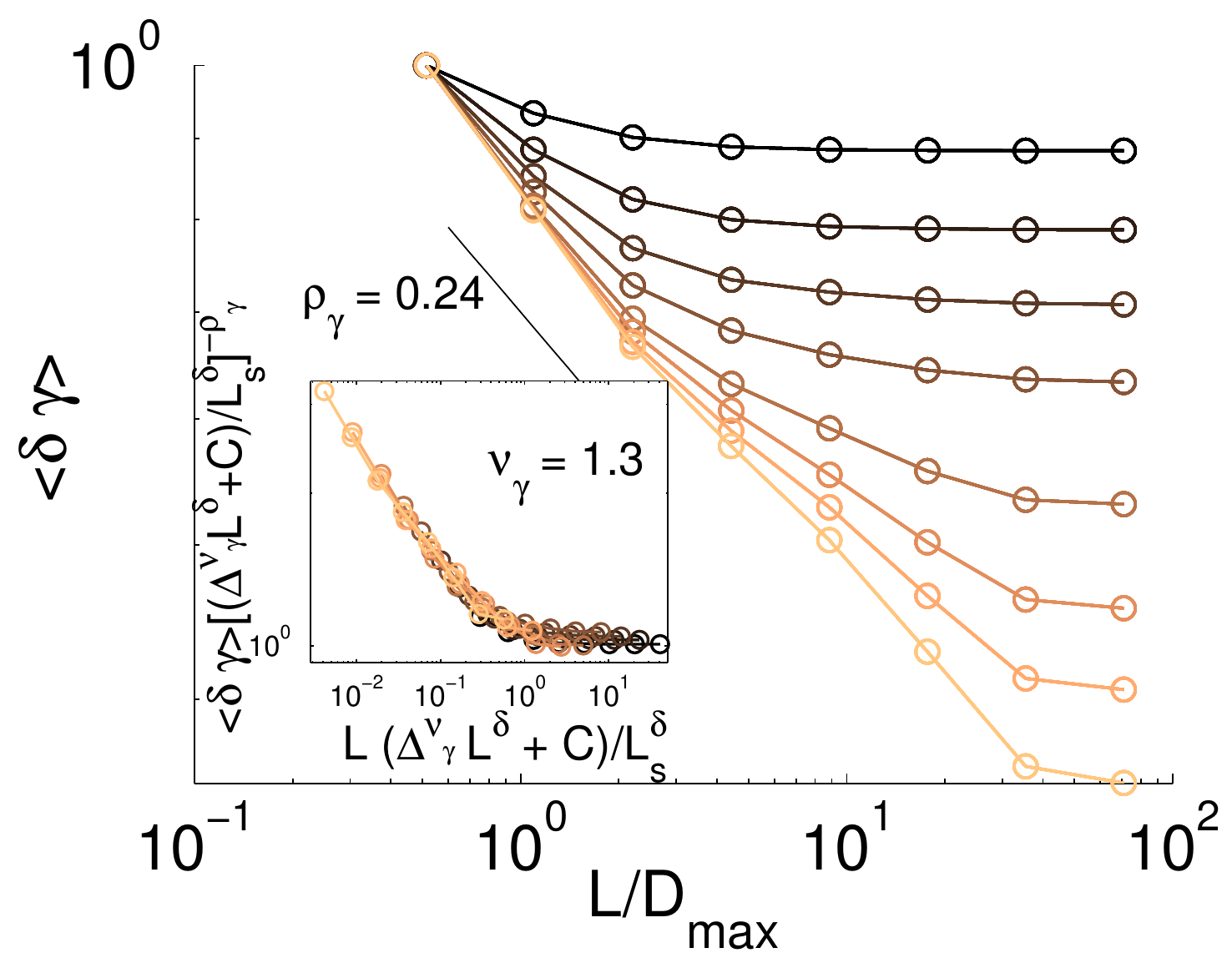}
 \end{minipage}
 \begin{minipage}{0.18\linewidth}
   \includegraphics[width=1\textwidth]{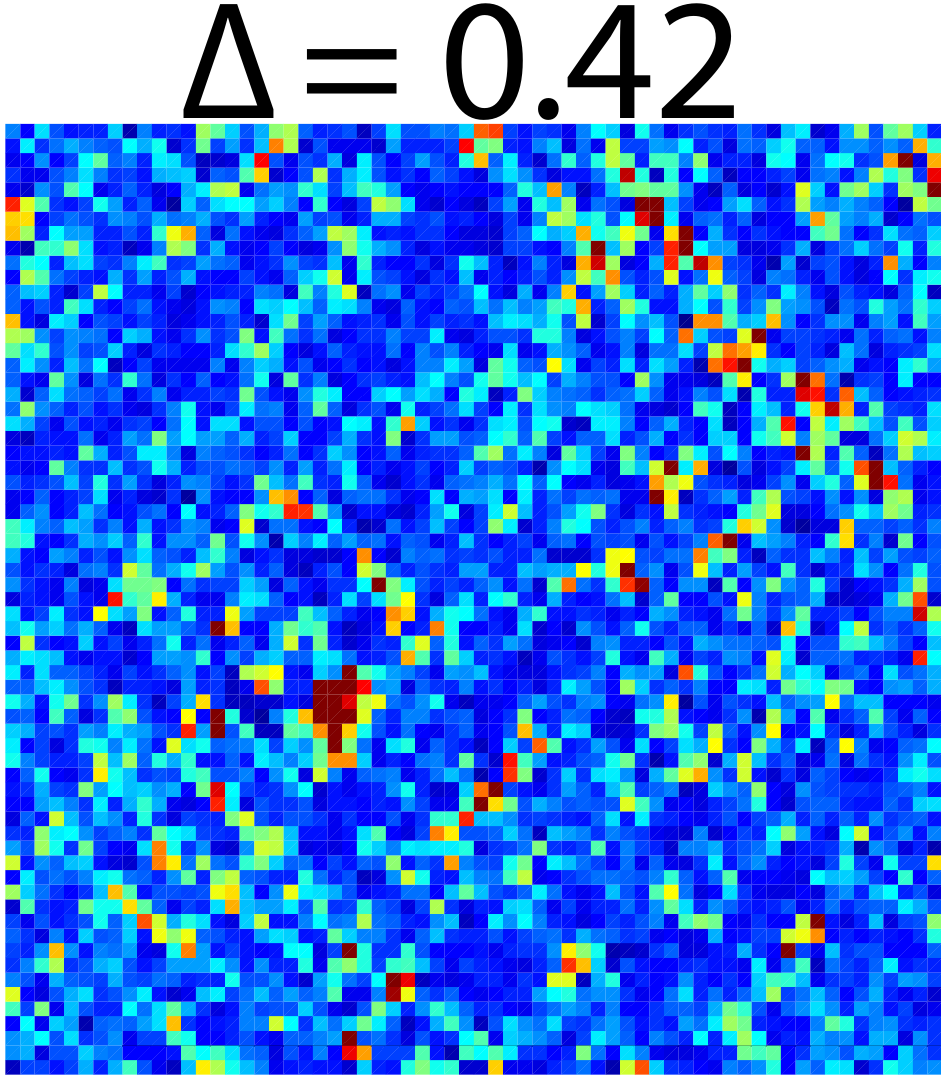}
   \includegraphics[width=1\textwidth]{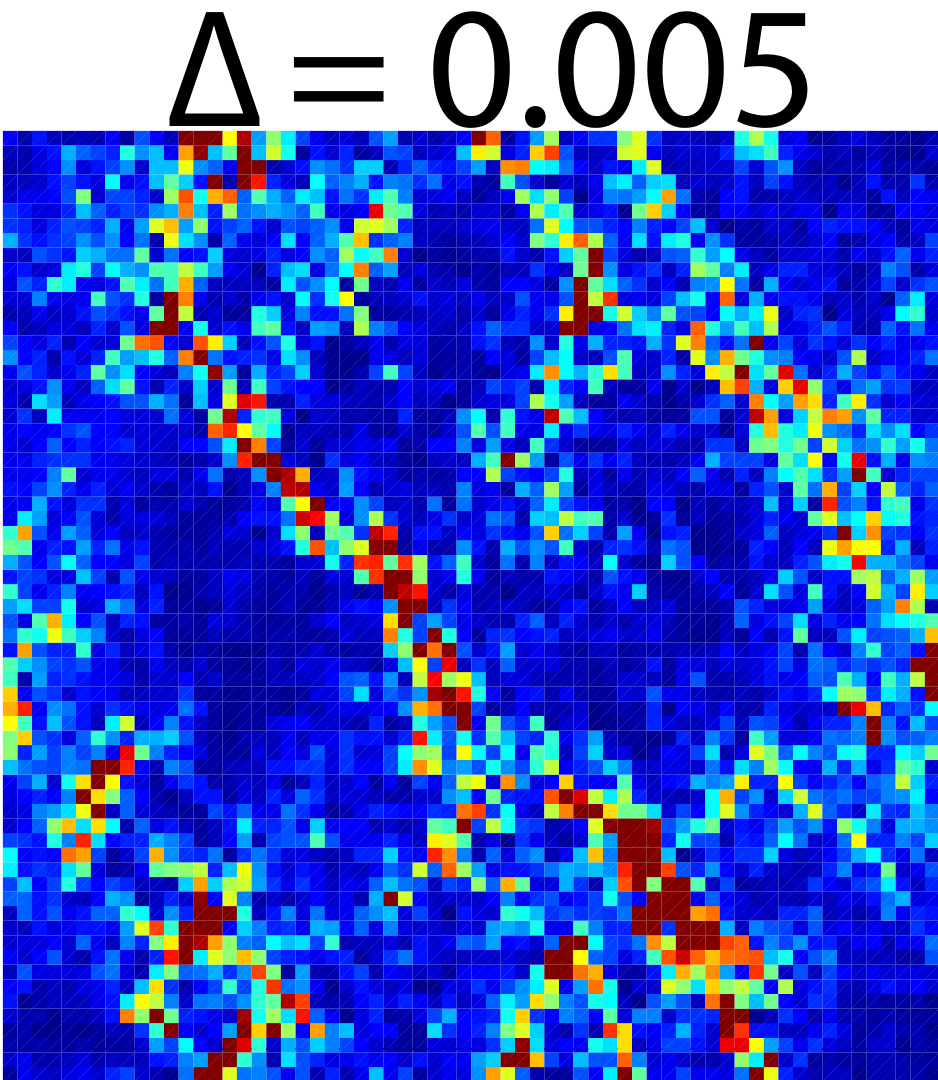}   
 \end{minipage}
 \hrule
 \begin{minipage}{0.15\linewidth}
 \includegraphics[width=1\textwidth]{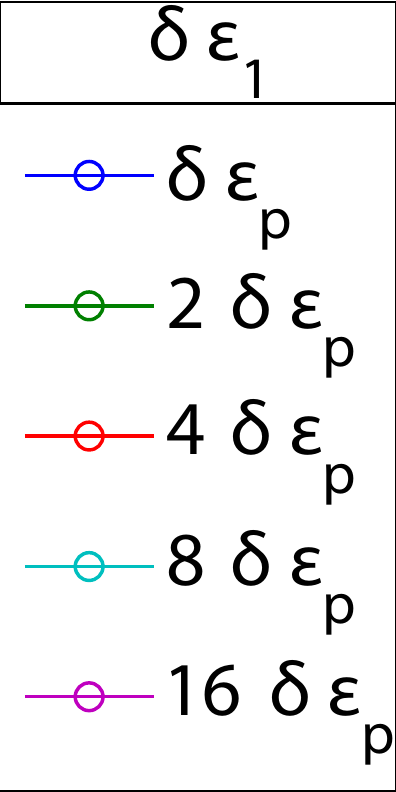}
\end{minipage}
 \begin{minipage}{0.55\linewidth}
 \includegraphics[width=1\textwidth]{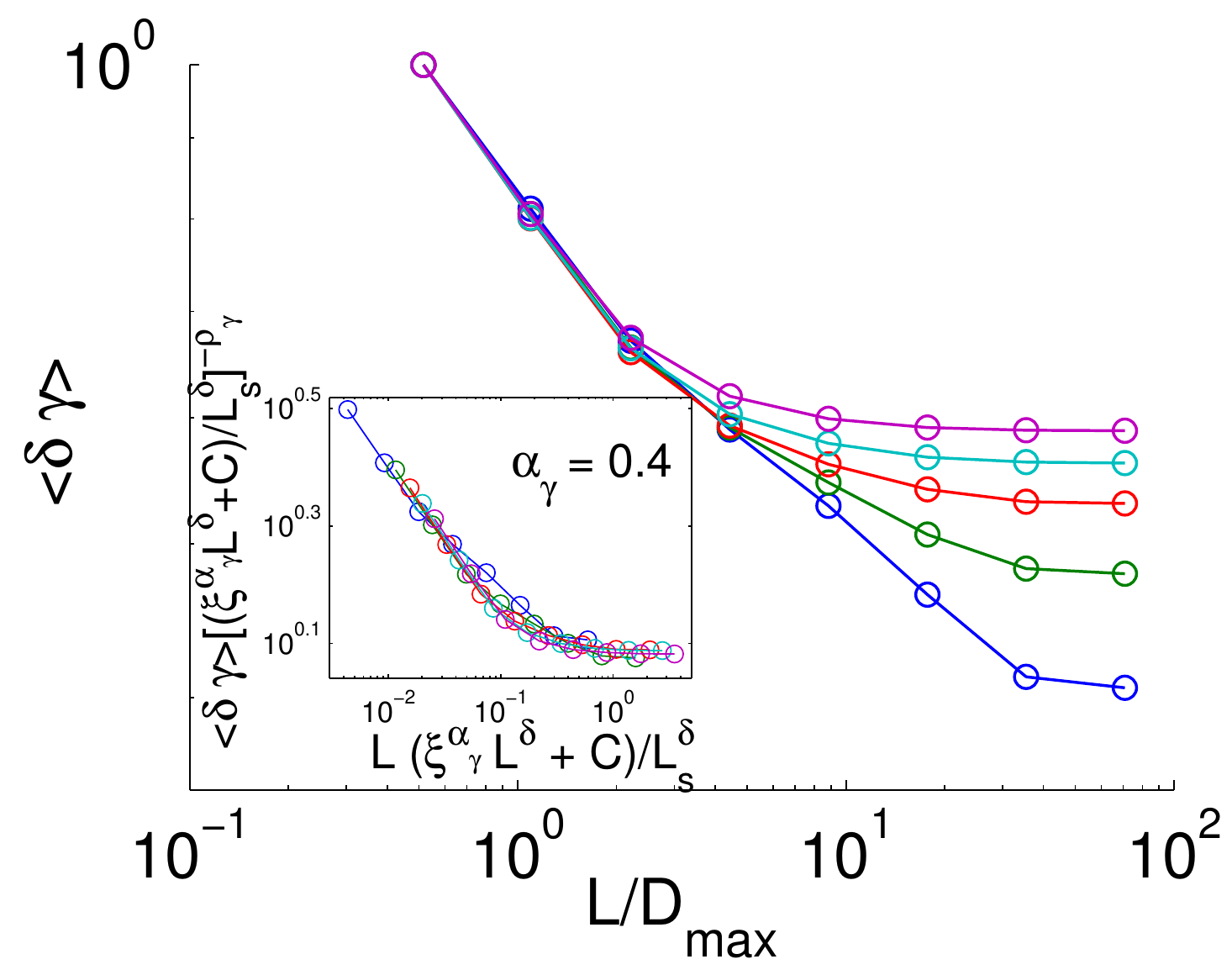}
 \end{minipage}
 \begin{minipage}{0.18\linewidth}
   \includegraphics[width=1\textwidth]{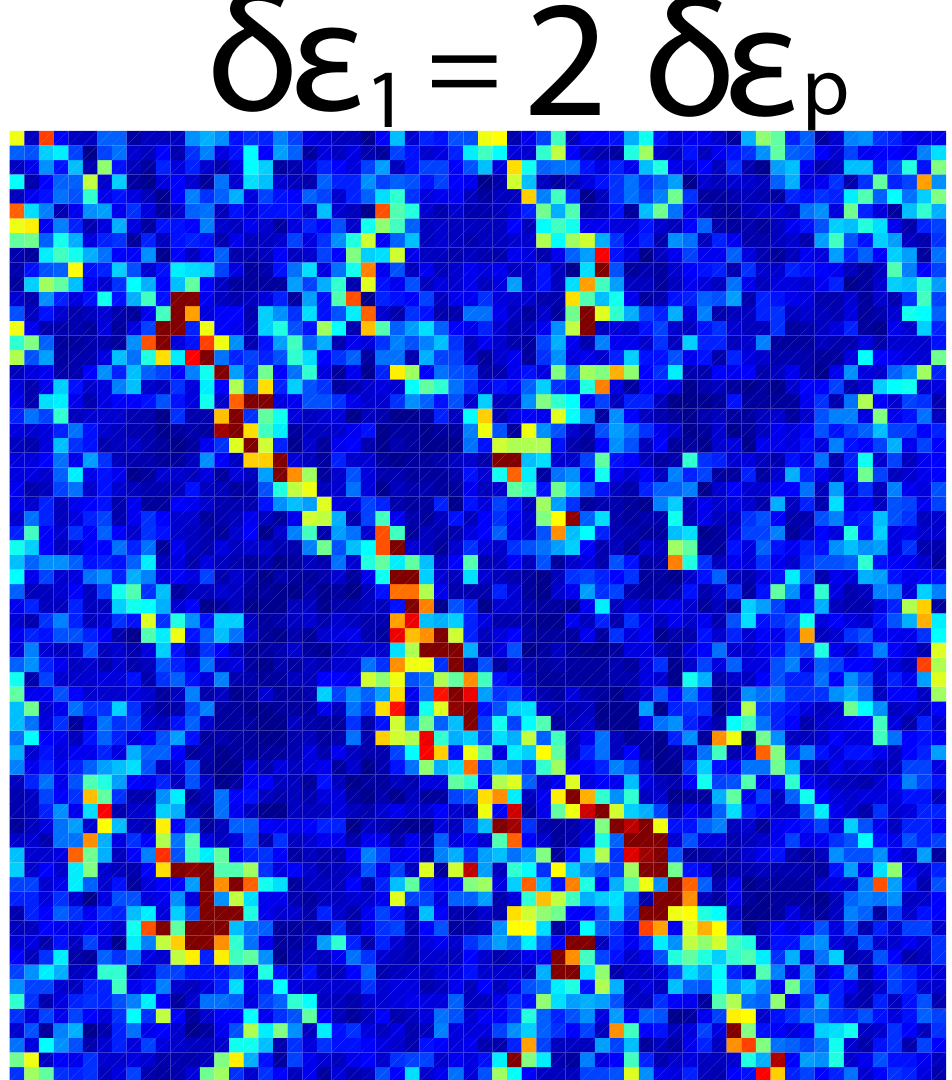}
   \includegraphics[width=1\textwidth]{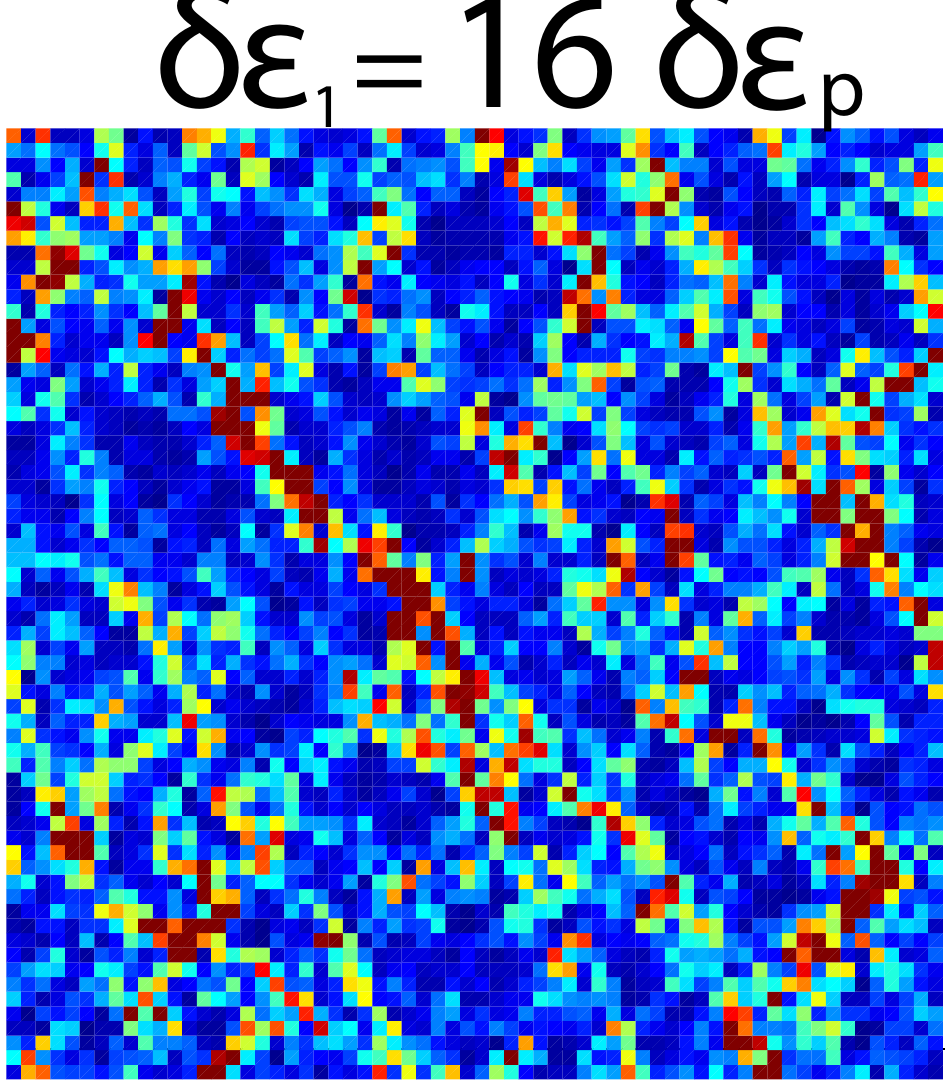}   
 \end{minipage}
\caption{Multi-scale analysis performed on the incremental shear strain field $\delta \gamma$. A selection of corresponding fields is shown on the right side: an arbitrary color scale (not shown) has been chosen for each snapshot. Top:~$<\delta{\gamma}>$ versus $L$ for decreasing values of $\Delta$ towards the critical point. The deformation scale $\delta \epsilon_1 = \delta \epsilon_p = 1.10^{-5}$ is used to compute values of $<\delta{\gamma}>$. 
The inset displays data collapse with respect to $\Delta$ (equation~\ref{eq::Collapse}). We find $\nu_\gamma = 1.3$, $C = 0.5$ and $\delta = 1$. Bottom:~$<\delta{\gamma}>$ versus $L$ at $\Delta = 0.005$ when increasing $\delta \epsilon_1$ from $\delta \epsilon_1 = \delta \epsilon_p$ to $\delta \epsilon_1 = 32 \delta \epsilon_p$. The inset displays data collapse with respect to $\xi = \frac{\delta \epsilon_1 - \delta \epsilon_p}{\delta \epsilon_p}$ (equation~\ref{eq::3}). We find $\alpha_\gamma = 0.4$. For graphical convenience, all computed values have been normalized by the ones computed at the micro-scale.\label{fig::StrainScaleAnal}}
\end{figure}
To study whether similar observations can be reported on the shear strain field, we consider a delaunay triangulation performed on the grain centers (Figure~\ref{fig::CoarseGrainingMethod}), after having removed the rattlers grains from the grain set. Then, we compute the partial derivatives at the mesh scale as $\epsilon_{ij} = 1/2 (\partial u_i/\partial x_j + \partial u_j/\partial x_i)$, where $(u_1,u_2)$ and $(x_1,x_2)$ are respectively the incremental displacements and spatial coordinates of grain centers. The coarse graining analysis is performed similarly than previously for stresses, here by averaging partial derivatives at corresponding spatial scales. An average shear strain rate~\cite{Invarients} is thus obtained as a function of $L$.
While not shown here, 
if one uses the constant timescale $T$ to compute incremental displacements, the correlation length associated with the shear strain rate field does not diverge as approaching the transition to the dense flow regime. Thus, the structure of the total strain field does not form simulteanously in time with the stress field. Intuitively, this would be the case if one would consider only the elastic component of the strain. Here, for $N_g=10000$, $\sim10^{4}$ stress increments are prescribed during the propagation time of an elastic wave throughout the sample. Hence, a multitude of contacts, in our case about 5\% of the whole contact network, are then sliding, although elastic interactions did not have time to travel across the entire sample.
Despite this, a progressive structuring of the shear strain field is observed when considering constant macroscopic deformation windows $\delta \epsilon_1=\delta\epsilon_p=1.10^{-5}$ to compute the scaling of $<\delta \gamma>$ (Figure~\ref{fig::StrainScaleAnal}). 
A divergence of the correlation length $l_\gamma^*$ similar to the one observed on the shear stress rate field is obtained as approaching the transition to the dense flow regime, as we find $\nu_\gamma=\nu_\tau=1.3$ from a collapse analysis (equation~\ref{eq::Collapse}). When considering larger macroscopic deformation windows $\delta \epsilon_1>\delta\epsilon_p$, the multi-scale properties of the deformation field are no longer observed.  This observation is in agreement with the shrinkage of the distributions of the fluctuation velocities at increasing timescales observed by~\cite{PRLRajai2002}. In this study, $\delta\epsilon_1=\delta\epsilon_p$ is the characteristic deformation value we need to consider in order to observe on the deformation field the critical behaviour already reported previously on the stress field at the timescale $T$ of an elastic wave propagation.
As the material softens when $\tau$ increases, the corresponding time of integration at constant deformation window $\delta \epsilon_1=\delta \epsilon_p$  decreases as the critical point is approached. This time can either be smaller or larger than the elastic wave traveling time $T$, depending on the imposed loading rate $\delta \sigma_1^{t_r}$. 
However, whatever the loading rate considered, $\delta \epsilon_1=\delta \epsilon_p$ remains equal to $1.10^{-5}$, pointing out that  
a given amount of plastic activity has to operate in order to observe multiscale properties within the incremental shear strain field. As the correlation lengths $l^*_\tau$ and $l^*_\gamma$ diverge the same way, the progressive structuring of the strain field is probably directly related to the progressive structuring of the stress field. 

All the presented scalings are undifferently obtained whatever the loading rate $\delta \sigma_1^{t_r}$ considered as soon as it is smaller than $\delta \sigma_1^{t_r}=1.10^{-6}$. Considering larger values for $\delta \sigma_1^{t_r}$, i.e. for example $\delta \sigma_1^{t_r}=5.10^{-6}$, no specific point materialized by the divergence of the correlation length can be reported~\cite{SuppMat}. Indeed, the mechanical behaviour of the granular assembly is in that case related to a dense flow regime from the initial stages of deformation, proving that the scalings observed here are associated to the transition from a quasi-static to a dense flow regime of deformation. 


\begin{figure}
\centering
 \includegraphics[width=0.3\textwidth]{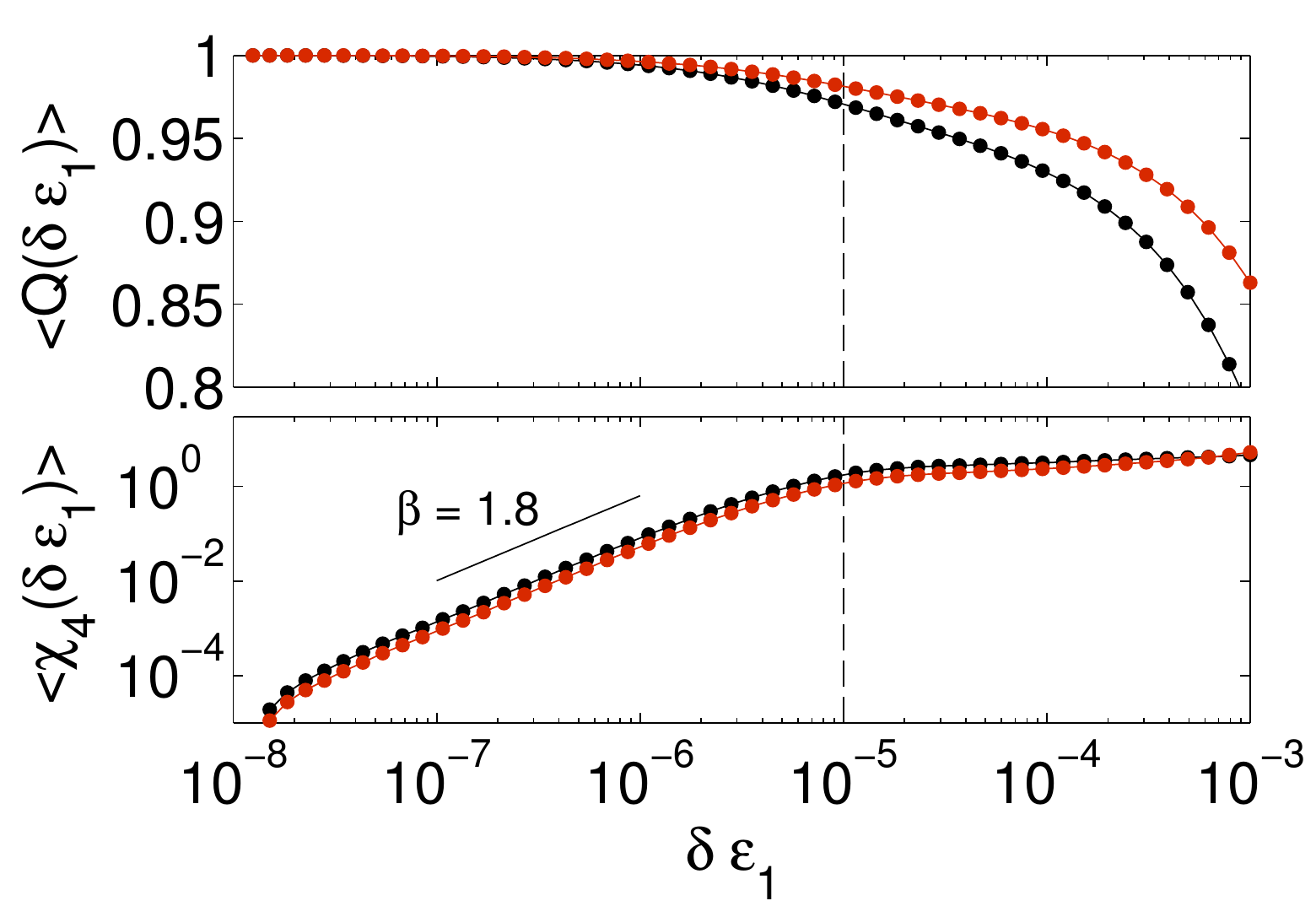}
 \caption{Susceptibility analysis performed in the quasi-static region on the sliding contacts belonging to the major (red curves) and minor (black curves) network. 
The vertical dashed line indicates the deformation value $\delta\epsilon_p=1.10^{-5}$. 
 \label{fig::SusceptSliding}}
 \end{figure}
An understanding of the characteristic value $\delta\epsilon_p$ 
can be obtained using a four-point dynamic susceptibility $\chi_4$~\cite{Abate2007Susceptibility,Aaron2007NaturePhys} analysis on the inter-particles contact network. 
From a contact configuration that we refer as ``initial'', selected at a value of axial deformation denoted $\epsilon_1^{init}$, we compute the self-overlap order parameter $Q_{\epsilon_1^{init}}(\delta \epsilon_1)=\frac{1}{N_c}\sum_{i=1}^{N_c}w_i$, where $N_c$ is the number of contacts that are not sliding in the initial configuration and $w_i$ is a step-function cutoff that equals $1$ if no sliding event has been recorded on contact $i$ over the whole deformation window $\epsilon_1^{init}~rightarrow~epsilon_1^{init}+\delta \epsilon_1$, and $0$ otherwise.
The first two moments $Q(\delta \epsilon_1)=< Q_{\epsilon_1^{init}}(\delta \epsilon_1) >$ and $\chi_4(\delta \epsilon_1)=N_c\big{[}< Q_{\epsilon_1^{init}}(\delta \epsilon_1)^2 >-< Q_{\epsilon_1^{init}}(\delta \epsilon_1)>^2\big{]}$ of $Q_{\epsilon_1^{init}}(\delta \epsilon_1)$ (calculated from sample-to-sample fluctuations) are then computed in the quasi-static region ($\tau < \tau_c$). Doing this, we evaluate from an initial configuration the number and the associated spatial heterogeneity of sliding events nucleation as axial deformation increases. 
Figure~\ref{fig::SusceptSliding} shows $<Q(\delta \epsilon_1)>$ and $<\chi_4(\delta \epsilon_1)>$, where $<.>$ here means an average over all the values of $\Delta$ (since no significant variation of $Q(\delta \epsilon_1)$ and $\chi_4(\delta \epsilon_1)$ is observed in the quasi-static region) computed by considering separately the major and minor force networks. The major force network is defined by selecting contact forces larger than the average. 
By construction, $<Q(\delta \epsilon_1)>$ is initialy equal to $1$. As $\delta \epsilon_1$ increases, $<Q(\delta \epsilon_1)>$ decreases but never reaches 0, meaning that a considerable amount of contacts never slide. About 35\% (respectively 55\%) of the contacts of the minor (respectively major) network did not slide at the end of the test, meaning that the permanent deformation is extremely localized and that rigid bodies remain throughout the whole test~\cite{Szarf2011}. The value of $<\chi_4(\delta \epsilon_1)>$ indicates, with respect to increasing axial deformation, the variability in the nucleation of new contact slidings. At low values of $\delta \epsilon_1$, it increases as $<\chi_4(\delta \epsilon_1)> \sim \delta \epsilon_1^\beta$ with $\beta = 1.8$, meaning that spatially correlated sites of contact sliding events are nucleating. As $\delta \epsilon_1$ exceeds the threshold value $\delta \epsilon_p = 1.10^{-5}$, $<\chi_4(\delta \epsilon_1)>$ saturates, as all the spatially correlated contacts located close to the coulomb criteria, i.e. susceptible to slide, have been destabilized. At this stage, only 4\% (respectively 3\%) of contacts have slided at least one time in the minor (respectively major) network. 

To conclude, in dense granular assemblies, incremental stress and strain fields are both characterized by a growing correlation length that diverges as approaching the onset of macroscopic instability, which can therefore be identified as a critical point. A similar behavior has been reported in compressive failure of continuous materials~\cite{JStatGirard2010,PRLGirard2012}. We interpret these stress and strain specific structures as resulting from dynamic stress redistributions induced by the local dissipation of elastic energy materialized by contact slidings. At macroscopic instability, a local contact sliding event induces correlated elastic stress perturbations up to the scale of the whole granular assembly. These features can only be observed when carefully examining characteristic timescales for stresses, and characteristic macroscopic strain increments for strains. These characteristic timescales may drastically be affected when considering different inital packing properties of the granular assemblies, e.g. considering low coordinated and/or loose inital samples, which has not yet been investigated in the present study. 






The last question that arises is to whether a limit in decreasing correlation length $l^*_\gamma$ on the shear strain field is reached at $\Delta \rightarrow 0$ for values of $\delta \epsilon_1$ much larger than $\delta \epsilon_p$, which would characterize the thickness of a perennial macroscopic shear band potentially formed at the onset of instability.
To investigate this, we hypothesize that, close to the critical point ($\Delta~\rightarrow~0$), $l^*$ varies as
\begin{equation}
l^*_\gamma\sim\frac{L_s^{\delta}}{L_s^{\delta}\xi^{\alpha_\gamma} + C}
\label{eq::3}
\end{equation}
where $\xi~=~\frac{\delta \epsilon_1~-~\delta \epsilon_p}{\delta \epsilon_p}$ and $\alpha_\gamma$ is the exponent of divergence with respect to $\xi$. This hypothesis is tested from a collapse analysis (inset of Figure~\ref{fig::StrainScaleAnal}). We find $\alpha_\gamma = 0.4$. This shows that $l^*_\gamma$ keeps decreasing as the considered deformation window size $\delta \epsilon_1$ is increased, showing that the correlation length only depends on the value of $\delta \epsilon_1$ and that no intrinsic scale of saturation, potentially associated to a shear band thickness, can be identified at the onset of macroscopic instability.

\acknowledgments
We thank Ga\"el Combe for having provided the discrete element model and for fruitfull discussions. We thank Jean Braun for having provided efficient routines to compute Voronoi tesselations.
All computations were performed at SCCI-CIMENT Grenoble.







\begin{thebibliography}{10}

\bibitem{Cates1998}
\Name{M.E.~Cates, J.P.~Wittmer, J.P.~Bouchaud, and P.~Claudin}
\REVIEW{Phys. Rev. Lett.}{81}{1998}{9}.

\bibitem{NatureLiu1998}
\Name{A.~Liu and S.~Nagel}
\REVIEW{Nature}{396}{1998}{21}.

\bibitem{Combe2000}
\Name{G.~Combe and {J-N.} Roux}
\REVIEW{Phys. Rev. Lett.}{85}{2000}{3628}.

\bibitem{NatureBehringer2011}
\Name{D.~Bi, J.~Zhang, B.~Chakraborty, and R.P. Berhinger}
\REVIEW{Nature}{480}{2011}{355}.

\bibitem{CRPhysiquePouliquen2002}
\Name{O.~Pouliquen and F.~Chevoir}
\REVIEW{C.R. Physique}{3}{2002}{163}.


\bibitem{Mogi1967}
\Name{K.~Mogi}
\REVIEW{J. Geoph. Res.}{72}{1967}{5117-5131}.

\bibitem{Haimson2000}
\Name{B.~Haimson and C.~Chang}
\REVIEW{Int. J. Rock Mech. Min. Sciences}{17}{2000}{285-296}.

\bibitem{Viggiani2004}
\Name{J.~Desrues and G.~Viggiani}
\REVIEW{Int. J. Numer. Anal. Meth. Geomech.}{28}{2004}{279}.

\bibitem{Hall2010}
\Name{{S.~A.}~Hall, {D.~M.} Wood, {E.} Ibraim, and G. Viggiani}
\REVIEW{Granular Matt.}{12}{2010}{1-14}.

\bibitem{RudnickiRice1975}
\Name{J.W. Rudnicki and J.R. Rice}
\REVIEW{J. Mech. Phys. Solids}{23}{1975}{371}.

\bibitem{HaimsonRudnicki2010}
\Name{B.C. Haimson and J.W. Rudnicki}
\REVIEW{J. Struct. Geol.}{32}{2010}{1701}.

\bibitem{Tordesillas2007}
\Name{A.~Tordesillas}
\REVIEW{Phil. Mag.}{87}{2007}{32}.

\bibitem{PRLRajai2002}
\Name{F.~Radjai and S.~Roux}
\REVIEW{Phys. Rev. Lett.}{89}{2002}{064302}.

\bibitem{BookRajaiDubois2011}
\Name{F.~Radjai and F.~Dubois}
\Book {Discrete Numerical Modeling of Granular Materials}
\Editor{Wiley-ISTE}
\Year{2011}.

\bibitem{SuppMat}
See supplementary material for details and extended results. 

\bibitem{Cundall1979}
\Name{P.A. Cundall and O.D.L. Strack}
\REVIEW{G\'eotechnique}{29}{1979}{47}.

\bibitem{Agnolin2007}
\Name{I.~Agnolin and J-N.~Roux}
\REVIEW{Phys. Rev. E.}{76}{2007}{6}.

\bibitem{Brocher2008}
\Name{T.M. Brocher}
\REVIEW{BSSA}{98}{2008}{2}.

\bibitem{GDRMidi2005}
\Name{G.D.R. Midi}
\REVIEW{Eur. Phys. J. E.}{14}{2004}{341}.

\bibitem{Combe2003}
\Name{G.~Combe and {J-N.} Roux}
\REVIEW{Deformation Charactersitics of Geomaterials, 3ème Symposium sur
  le Comportement des sols et des roches tendres, Lyon, 22-24 Septembre 2003,
  In di Benedetto et al.}{}{2003}{1071}.


\bibitem{PRLMarsan2004}
\Name{D.~Marsan, H.~Stern, R.~Lindsay, and J.~Weiss}
\REVIEW{Phys. Rev. Lett.}{93}{2004}{178501}.

\bibitem{JStatGirard2010}
\Name{L.~Girard, D.~Amitrano, and J.~Weiss}
\REVIEW{J. Stat. Mech.}{}{2010}{P01013}.

\bibitem{Invarients}
the invariant $x$, where $x$ represents $\tau$ for the shear stress field or
  $\gamma$ for the shear strain field, is computed as $x = x_{I} - x_{II}$,
  where $x_{I}$ and $x_{II}$ are the principal components of the stress
  or strain tensor.

\bibitem{DeGiuliMcElwaine2011}
\Name{E.~DeGiuli and J.~McElwaine}
\REVIEW{Phys. Rev. E}{84}{2011}{041310}.

\bibitem{Abate2007Susceptibility}
\Name{A.~R. Abate and D.~J. Durian}
\REVIEW{Phys. Rev. E}{76}{2007}{021306}.

\bibitem{Aaron2007NaturePhys}
\Name{A.~S. Keys, A.~R. Abate, S.~C. Glotzer, and D.~J. Durian}
\REVIEW{Nature Phys.}{3}{2007}{260}.

\bibitem{Szarf2011}
\Name{K.~Szarf, G.~Combe, and P.~Villard}
\REVIEW{Powder Tech.}{208}{2011}{239}.

\bibitem{PRLGirard2012}
\Name{L.~Girard, J.~Weiss, and D.~Amitrano}
\REVIEW{Phys. Rev. Lett.}{108}{2012}{225502}.

\end{thebibliography}
\end{document}